\documentclass[%
showkeys,
preprint,
amsmath,amssymb,
aps,
pra,
]{revtex4-1}
\usepackage{graphics}
\usepackage{epsfig}
\usepackage{epstopdf}
\usepackage{amsmath}
\usepackage{array}
\usepackage{subfigure}
\usepackage[colorlinks=true,linkcolor=blue,citecolor=blue,      urlcolor=blue]{hyperref}
\begin{document}

\title{Emergence of space and expansion of universe}
\author{Hassan Basari V. T. $^1$}
\email{basari@cusat.ac.in}
\author{P. B. Krishna $^1$}
\email{krishnapb@cusat.ac.in}
\author{Priyesh K. V $^2$}
\email{priyeshkv@stpauls.ac.in}
\author{Titus K. Mathew. $^1$}
\email{titus@cusat.ac.in}
	\affiliation{%
	$^1$ Department of Physics, Cochin University of Science and Technology, Kochi, Kerala 682022, India.}
\affiliation{%
	$^2$ Department of Physics, St. Paul's College, Kalamasseri, Kochi-03, India
}%

\begin{abstract}
According to the principle of emergence, the expansion of the universe can be explained as the emergence of space with the progress of cosmic time. We have analytically solved the equation of emergence proposed by Padmanabhan by assuming the Komar energy density $\rho+3P$ as a function of the Hubble parameter. The resulting model describes the evolution of the universe, which proceeds towards a final de Sitter state. Model parameters have been extracted using the cosmological observational data. Further, the horizon entropy evolution of the model has been studied.  The model predicts a universe having a transition from a prior decelerated epoch to a late accelerated epoch and reasonably predicts the cosmological constant. 
\end{abstract}
\maketitle
%
%
%
%
%

\section{Introduction}
\label{Intro}

Recent studies indicate a deep connection between gravity and thermodynamics\cite{AKBAR20067,Cai_Kim,paddy2010dec,paddynew3,paddynew4,paddynew1,paddynew2,sheykhi2007,disformalfr2019}. It emerged first from the blackhole thermodynamics, established in the 1970s by Bekenstein, Bardeen, and Hawking\cite{bekenstein1,bardeen,bekenstein2,Hawking1,Hawking2}. Later, Jacobson\cite{1995} showed that 
Einstein's field equations could be derived from Clausius relation on a local Rindler causal horizon using the horizon area-entropy relation\cite{bekenstein1}. In 
2010, Padmanabhan\cite{paddy2010dec} derived Newton's law of gravity by combining the equipartition law of energy at the horizon and the thermodynamic relation $S = \frac{E}{2T}$ where 
$S$ is the entropy, $E$ is the effective gravitational mass, and $T$ is the temperature of the horizon. Verlinde reformulated gravity as an entropic force that arose from the natural tendency of material distribution to maximise the entropy, and he derived Newton's law of gravitation and the more general Einstein's field equations\cite{Verlinde}. Further, it has been shown that gravitational field equations in Einstein's gravity and more general theories of gravity can yield themselves to a thermodynamical interpretation\cite{paddynew3,paddynew4,Kothawala2007,Kothawalanew1}. In references \cite{paddynew1,paddynew2}, authors have shown that gravitational equations of motion can be obtained from thermodynamical extremum principles. All these lead to the intriguing idea that gravity could be an emergent phenomenon, as is the case with thermodynamics.

In the emergent gravity paradigm, it is argued that field equations are emergent while assuming the pre-existence of space-time. Here the gravitational field equations emerge as the thermodynamic limit of the underlying statistical mechanics of microscopic degrees of freedom \cite{paddynew3,Kothawala2007, Chakraborty2015,Chakraborty2017,Paddy:2017intgratingconstant}. 
Motivated from this, Padmanabhan took one step further to state that 
expansion of the universe can be considered as the emergence of space with the progress of cosmic time.  It is not easy to treat time as being an emergent entity. However, there is a unique time in cosmology, the proper time for all geodesic observers for whom the cosmic background radiation is homogeneous and isotropic.  
Following this, it was proposed that the expansion of the universe, or more correctly, the emergence of space,  
is driven by the discrepancy between the degrees of freedom (DoF) on the horizon and that of the bulk enclosed
by the horizon \cite{paddy2012jun}. It can be described in a mathematical form, 
$\frac{dV}{dt} = L_{P}^{2} \left(N_{sur} - \epsilon N_{bulk}\right),$ where $N_{sur}$ is the 
DoF on Hubble horizon, $N_{bulk}$ is that in bulk within the horizon, $V$ is the Hubble volume, $L_{P}^{2}$ is the Planck area, and $\epsilon$ is a factor incorporated to make the $N_{bulk}$ a positive definite quantity. As the cosmic time progresses or as the universe expands, $N_{bulk}$ increases and approaches $N_{sur}$ (meantime $N_{sur}$ also increases with time) and finally achieves the condition $N_{bulk}=N_{sur},$ known as the holographic equipartition at which 
the universe enter the
de Sitter epoch of constant Hubble volume.
The viability of this principle relies on the fact that it is possible to derive
Friedmann's equations from it,
for Einstein's gravity\cite{paddy2012jun,cai} and more general gravity theories
like Gauss-Bonnet 
and 
Lovelock gravity \cite{cai,Sheykhi2013}. This principle also leads to the understanding of a cosmological constant\cite{hamsa,paddycosmologicalconstant,paddyhamsacosmin,krishnampla}. In references\cite{FLDezaki,2018}, authors have shown that the principle of emergence proposed by Padmanabhan can be obtained directly from the fundamental laws of thermodynamics, which further strengthen the viability of the principle\cite{Hareesh_2019}. Recent studies show that the expansion law effectively implies the horizon entropy maximisation in Einstein's gravity \cite{krishna1}, and in more general forms of gravity like Gauss-Bonnet and Lovelock gravities \cite{krishna2}.

In the present work, we derive a cosmological model from the principle of emergence proposed by Padmanabhan, by obtaining an analytical solution for the Hubble parameter. We found that the model very well explains the late accelerating universe and predicts a reasonable value for the cosmological constant.

The paper is organised as follows. In section \ref{sec.2}, we present the derivation of the solution to the principle of the emergence of space by taking a suitable form for the gravitational energy density, the Komar energy density, and show that the corresponding model of the universe predicts a transition from a prior decelerated epoch to the late accelerated stage. We extracted the model parameters by using observational data, then analyse the entropy evolution in the model. In section  \ref{sec:4},  we present how exactly the model can be interpreted as the emergence of space. The conclusions are presented in section \ref{sec:5}.

\section{Possible solution to the principle of emergence and its admissibility}
\label{sec.2}

According to Padmanabhan, the expansion of the universe is driven by the quest of the universe towards the  holographic equipartition, $ N_{sur}= N_{bulk},$ and the principle can be expressed as 
\begin{equation}\label{eq:law}
\frac{dV}{dt} = L_{P}^{2} \left( N_{sur} -\epsilon N_{bulk} \right),
\end{equation}
where the Hubble volume, $ V = 4\pi/3H^3,$ the Hubble volume.   
The degrees of
freedom on the horizon   
and that within bulk enclosed by the horizon
are defined as \cite{paddy2012jun}
\begin{equation}
\label{eqn:DoF}
N_{sur} = \frac{4\pi}{L_{P}^{2}H^{2}}, \hspace{0.6in}   N_{bulk} = \frac{\mid E \mid}{\frac{1}{2}k_B T} = -\epsilon \frac{\left(  \rho + 3P \right)  V}{\frac{1}{2}{k_BT}}\, ,
\end{equation}
where $ T = \frac{H}{2\pi}$ is the Hawking's temperature, $(\rho+3P)V$ is the Komar energy and
$\epsilon$ is defined as
\begin{equation}
\label{eqn:epsilon}
\epsilon = +1~ \textrm{for} \left( \rho + 3P \right) < 0 ~~ \textrm{and}~~ \epsilon = -1 ~ \textrm{for} \left( \rho + 3P \right) > 0.
\end{equation}
Here $N_{bulk}$ is the degrees of freedom within the horizon which are in equipartition with the horizon at temperature $T.$ 

Using equations(\ref{eqn:DoF}) and (\ref{eqn:epsilon}),
equation  
(\ref{eq:law})
can be rewritten as
\begin{equation}\label{eqn:lawexpanded}
-\frac{4\pi\dot{H}}{H^4} = L_{P}^{2}\left[  \frac{4\pi}{L_{P}^{2}H^{2}} + \frac{16\pi^{2}\left( \rho + 3P\right)  }{3H^{4}} \right] ,
\end{equation}
where $\dot H$ is the derivative of $H$ with cosmic time $t.$
To find the evolution of the Hubble parameter, it is necessary to substitute the Komar energy density, $\rho+3P.$ 

The Komar energy $E$ of a gravitating system can be determined using the relation \cite{Padmanabhan:2003Grentropy},
\begin{equation}
E = \int_{\nu} d^3x \quad 2\sqrt{\gamma} N \left(T_{ab}-\frac{1}{2}g_{ab}T\right) u^au^b ,
\end{equation}
where $\gamma$ is the induced metric on the three space and 
$2\left(T_{ab}-\frac{1}{2}g_{ab}T\right)u^au^b$ reduces to the Komar energy 
density, $\left(\rho+3P\right),$ with $u^a$ being the four-velocity. For a region bounded by the Hubble sphere of volume, $V,$ the above integral reduces to, $E=(\rho+3P) V.$  This total energy can also be expressed in terms of the horizon entropy, $S$ as, $2TS$ 
\cite{Padmanabhan:2003Grentropy,Padmanabhan2014}. 
For an expanding universe the horizon area-entropy relation implies that, 
$S \propto \frac{1}{H^2}$. Together with the relation
$T \propto H$ \cite{paddy2012jun}, it is not hard to deduce that
the Komar energy $E$ will be proportional to $ H^{-1}.$ 
Having the 
volume of the horizon as $V \sim \frac{1}{H^3}$ 
it can be deduced that 
Komar energy density $\rho+3P$ is proportional to $H^{2}$ in the equilibrium description. However, in general, the Komar energy density can also depends on $\dot{H}$  in the dynamic situations. 
Taking account of these facts, we will have a general form for Komar energy density as,

\begin{equation}\label{eqn:Komarnew}
\rho + 3 P = -B_2+ B_1 H^2 +B_3\dot{H}, 
\end{equation} 
where $B_{1}$, $B_{2}$ and $B_{3}$ are finite positive constants. This form for Komar energy density ensures that in the prior epoch in which $H$
is relatively large, $\rho+3P>0,$ while in the late epoch at which $H$ is sufficiently less, $\rho+3P<0$ will be
satisfied. The first negative constant term in the above equation is required to take care of the end de Sitter epoch.  This form  can also be obtained by expressing the Komar energy density as a Taylor expansion in terms of Hubble parameter H and avoiding the odd terms for ensuring the general covariance
\cite{runningvacuum,generalcovariance1,generalcovariance2,generalcovariance3,generalcovariance4}.

Now we substitute the Komar energy density given in equation (\ref{eqn:Komarnew})  into the equation (\ref{eqn:lawexpanded}), we have,  
\begin{equation}
\frac{-4\pi}{H^{4}}\dot{H} = L_{P}^{2}\left(\frac{4\pi}{L_{P}^{2}H^{2}}+\frac{\epsilon^{2} 16\pi^2\left(-B_{2} + B_{1}H^{2}+B_3\dot{H}\right)}{3H^{4}}\right),
\end{equation}
which can be reduced to the form
\begin{equation} \label{eq:newlaw}
\frac{dH}{dt} = \alpha^{2} - \beta^2 H^{2},
\end{equation}
where $\alpha = \frac{1}{\gamma}\sqrt{\frac{4\pi L_{P}^{2}B_{2}}{3}}$, $ \beta =\frac{1}{\gamma} \sqrt{ 1 + \frac{4\pi L_{P}^{2}B_{1}}{3}}$ with $ \gamma =\sqrt{ 1 + \frac{4\pi L_{P}^{2}B_{3}}{3}} $ are positive constants. Here, $\alpha$ has the dimension of $H$, while $\beta$ and $\gamma$ are dimensionless. To solve the expansion law, we took the condition, $\dot{H} \leq  0$ or equivalently $\frac{\alpha^2}{\beta^2} \leq H^{2}$. This condition ensure the decreasing nature of the Hubble parameter with the progress of cosmic time. The alternative solutions with $\dot{H}  > 0$ (phantom behaviour) lead to the breakdown of causality and the violation of the second law of thermodynamics\cite{krishna1,krishna2}, and it also leads to quantum instabilities \cite{phantomproblems1,phantomproblems2}.

The solution of equation (\ref{eq:newlaw}), following the admissibility condition, $\dot{H} <0,$ owing to fact that the $H$ decreases continuously with cosmic 
time is, 
\begin{equation}\label{eqn:Hm1}
H = \frac{\alpha\left( D_1e^{2\alpha\beta t}+1\right)}{\beta \left( D_1e^{2\alpha\beta t}-1\right)}
\end{equation}
and the corresponding scale factor is,
\begin{equation}\label{eqn:Ha1}
a= D_2 \left(\frac{D_1 e^{2\alpha\beta t} -1}{e^{\alpha\beta t} }\right)^{1/\beta^{2}} ,
\end{equation}
where $D_1$ and $D_{2}$ are the integration constants. 
The asymptotic behaviour of the above solutions
is as follows. 
In the limit $t\to 0$ the Hubble parameter tending to a very high value and scale factor $a \to 0$ provided $D_1 \approx 1$. 
While as $t\to \infty$ the Hubble parameter evolves to a constant, $H \to \alpha/\beta$ and the scale factor $a \to e^{(\alpha/\beta)t},$ which is equivalent to a de Sitter type behaviour.  
It thus seems that this 
solution 
predicts a transition into the late accelerated epoch. 

An important consequence of the above solution is that the Komar energy density in expression (\ref{eqn:Komarnew}) obeys the null energy condition, $\rho+P \ge 0.$ After obtaining the pressure $P$ from equation (\ref{eqn:Komarnew}),  the  the null energy condition can be expressed as, 
\begin{equation}\label{ineq1}
\rho + P = \frac{-B_{2}}{3} +\frac{B_{1}H^{2}}{3}+\frac{B_{3}\dot{H}}{3}+ \frac{H^{2}}{4\pi L_{P}^{2}} \ge 0.
\end{equation}
In formulating this equation, we have substituted the energy density $\rho,$ using the Friedmann equation, $H^2=(8\pi L_P^2/3) \rho.$
On substituting for the constants $B_1, B_2$ and $B_3$ in terms of the model parameters, $\alpha, \beta$ and $\gamma,$ we get,
\begin{equation}
-\alpha^2 \gamma^2 + \beta^2 \gamma^2 H^2 + \left(\gamma^2 -1\right) \dot H \geq 0.
\end{equation}
This inequality can be further simplified by using equation (\ref{eq:newlaw}, and as a result the null energy condition finally attain the simple form,
\begin{equation}
 H^2 \geq \frac{\alpha^2}{\beta^2}.
 \end{equation}
  This is the condition we have used to derive the solution given in (\ref{eqn:Hm1}) for an expanding universe. Thus it is evident that the assumed form of Komar energy density satisfies the null energy condition.

Now we will analyse the evolution of the the deceleration parameter, $q=-1-\dot H/H^2$ of the model. Using the Hubble parameter in equation (\ref{eqn:Hm1}), the deceleration parameter can be obtained as,
\begin{equation}
q = -1 +\frac{4\beta^2 e^{2\alpha\beta t}}{(1+e^{2\alpha\beta t})^2}.
\end{equation} 
As $t \to 0,$ parameter $q \to -1 + \beta^2$ indicate a prior decelerated phase, while as
$t \to \infty$, the parameter $q \to -1,$ which implies the late de Sitter
epoch. The prior decelerated epoch demands the condition, $\beta^2>1.$ We extract the parameter values in a later section, where it arises that, $\beta$ is indeed greater than one. We will now study the thermodynamic evolution 
corresponding to this solution in the next section. 

\subsection{ Entropy evolution and observational constraint}
It is known that any closed macroscopic system, like the universe, evolves to an equilibrium state of maximum entropy. In such case, the entropy $S$ of the system must satisfy \cite{Pavon2013},
\begin{equation}
\dot{S} \geq 0,\textrm{ always};\quad\ddot{S} < 0 \textrm{ at least in the final stage},
\end{equation}
where the over-dot represents the derivative with respect to the cosmic time. The first one is known as the generalised second law of thermodynamics, and the second is the convexity condition, which is to be satisfied in the asymptotic limit  $t \to \infty$ ( $a \to \infty.$). 

The total entropy of the universe can be approximated as the horizon entropy \cite{Egan,Pavon1}. From Bekenstein's result \cite{bekenstein1}, the entropy of the Hubble horizon can be expressed as
\begin{equation}
S_{H} = \frac{A_{H}}{4},
\end{equation}
where  $A_{H}$ is the area of the Hubble horizon with radius $\frac{c}{H}$. Substituting for $A_{H}$,
\begin{equation}\label{eq:entropy of horizon}
S_{H} = \frac{\pi c^{2}k_{B}}{L_{P}^{2}}\frac{1}{H^{2}}
=  \frac{K}{H^2},
\end{equation}
where $ L_{P} = \sqrt{\frac{\hbar G}{c^3}}$ is the Planck length and $K = \frac{\pi c^{2}k_{B}}{L_{P}^{2}}$. Now, we analyse the entropy evolution corresponding to the 
the solution given in 
equations (\ref{eqn:Hm1}).
In this case, 
we fix $D_{1} \approx 1$ subjected to the initial condition, $\lim\limits_{t \to 0}a(t) \approx 0$ following the relation (\ref{eqn:Ha1}). Using equation (\ref{eq:entropy of horizon}), the entropy of horizon 
	can be expressed as,
\begin{equation}
S=K\,\frac{\beta^{2}}{\alpha^{2}} \left( \frac{e^{2\alpha\beta t}-1}{ e^{2\alpha\beta t}+1}\right) ^2 .
\end{equation}
The rate of change of entropy with cosmic time can then be obtained as
\begin{equation}
\dot{S} = \frac{8K\beta^3}{\alpha} \frac{\left( e^{2\alpha\beta t}-1\right) e^{2\alpha\beta t}}{\left(   e^{2\alpha\beta t}+1\right)^3 }\,\,.
\end{equation}
It is clear that $\dot S>0$, for $\alpha>0$ and $\beta>0$, which 
indicate the validity of
the second law of thermodynamics. To check the maximisation condition of entropy we obtained $\ddot S$ as,
\begin{equation}
\ddot{S} = 16K\beta^4 e^{4\alpha\beta t} \frac{\left( - e^{2\alpha\beta t} + 4 - e^{-2\alpha\beta t}\right)} {\left(   e^{2\alpha\beta t}+1\right)^4 }.
\end{equation}
The quantity inside the parenthesis in the numerator controls
the evolution of
$\ddot S.$ As $t\to \infty$
the term $e^{-2\alpha\beta t}$ vanish and hence $\ddot S \to \frac{-16K\beta^4}{ e^{2\alpha\beta t}}$. This implies that the
$\ddot S$ approaches zero from below as $t\to \infty$,  ensuring the maximisation of entropy. So this solution is thermodynamically feasible. In this respect it should be noted the principle of emergence, can in fact lead to the maximisation of entropy as shown in references\cite{krishna1,krishna2}

Previously we had shown that this model admits a transition from a prior decelerated epoch to a late accelerated phase and a final de Sitter epoch.
The solution of the Hubble parameter, impose a constraint, $ H > \frac{\alpha}{\beta}$, which in turn constraints the age of the universe.
 In the present model, the age of the universe can
be expressed 
as,
\begin{equation}
t = \frac{1}{2\alpha \beta}ln\left(\frac{\frac{H\beta}{\alpha}+ 1}{\frac{H\beta}{\alpha}- 1}\right).
\end{equation}
This equation can be recast in to an appropriate form by multiplying it with $H$ as,
\begin{equation}
Ht = \frac{1}{\beta^{2}}\left[\frac{x}{2}ln\left(\frac{x+1}{x-1}\right)\right] = \frac{1}{\beta^{2}}y(x) ,
\end{equation}
where $x= \frac{H\beta}{\alpha},$ which is always greater than one and $y(x)$ is that within the square bracket of the above equation. It follows that the derivative of $y(x)$ satisfies
$y' < 0,$ which in turn indicates that $y(x)$ has a lower limit,
\begin{equation}
\qquad \lim_{x\to \infty}y = 1.
\end{equation}
From this it can be conclude that there exists a limit for $Ht$ as,
\begin{equation}
Ht \geq \frac{1}{\beta^2}.
\end{equation}
The above equation is valid for any $t>0.$ Hence the age of the universe, $t_0$ can be restricted as,
\begin{eqnarray}\label{eqn:ageconstrint}
H_0t_0 \geq \frac{1}{\beta^2},
\end{eqnarray}
where $H_0$ is the present value of the Hubble parameter.
This constraint to the age of the universe in the present model for which we need to determine the model parameters. 

\subsection{Extraction of the model parameters}

In this section,  
we estimate the model parameters, $(\alpha, \beta),$ and also the cosmological constant, $\Omega_{\Lambda} = (\alpha/\beta H_0)^2,$ by constraining the model 
	using the recent cosmological data, Supernovae Type Ia, observational Hubble data (H-Z data), BAO and CMB \cite{Pantheon1048data,hz52data,Shift2018,Blake2011}.
Here we adopt 
$\chi^{2}$- analysis method.
For Supernovae Type Ia data, we have used the Pantheon  
	data sets having 1048 data points. 
The magnitudes of supernovae 
at different redshifts $z_{i}$ 
can be computed using the relation,
\begin{equation}\label{eqn:mui}
\mu_{t}(\alpha, \beta, H_{0}, z_{i}) = 5\log_{10}\left[\frac{c(1+z_{i})}{Mpc}\int_{0}^{z_{i}}\frac{dz}{H(\alpha, \beta, H_{0}, z)}\right]+25 .
\end{equation}
For H-Z data we have used the 52 data points in the redshift range $z \in [0.07 - 2.36] $ \cite{hz52data}.

We used the shift parameter, $\mathcal{R}$ from the Planck 2018 observation \cite{Shift2018}. 
The shift parameter is related to the position of the first acoustic peak in the power spectrum of the cosmic microwave background (CMB) anisotropies, which can probe the high redshift region. The theoretical value of the shift parameter is defined as,\cite{cmb1997}
\begin{equation}\label{eqn:R}
\mathcal{R} =\sqrt{\Omega_{m}}\int_{0}^{z_{2}}\frac{dz}{h(z)},
\end{equation}
where $z_{2}$ is the redshift of the last scattering surface and $h(z)$ is the Hubble parameter corresponding to the 
redshift $z$ weighted with the present value of the Hubble parameter. 
\begin{figure}
	\centering
	\includegraphics[width=0.9\textwidth]{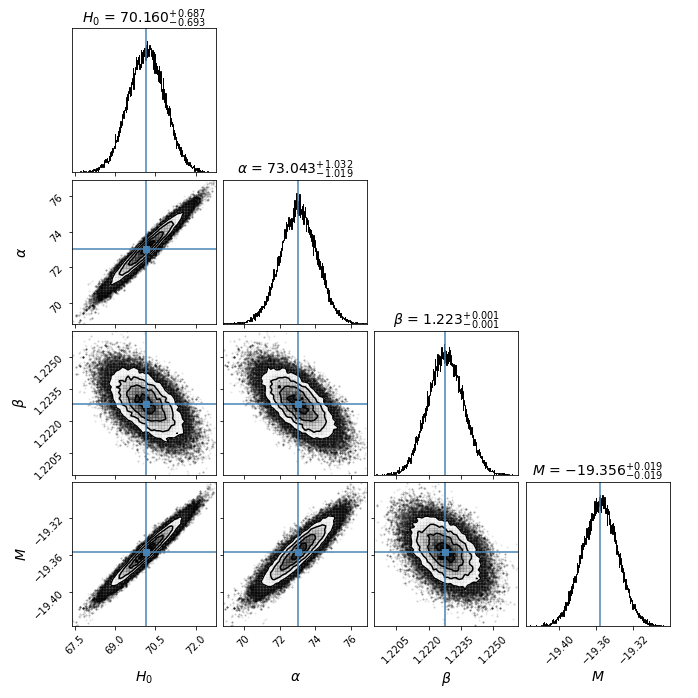}
	\caption{\label{fig:bestfit} Corner plots with marginal likelihood of model parameters  $ \alpha $, $  \beta $ , $H_0 $  from  SNe+CMB(Planck 2018)+BAO + HZ52 data sets.
		The best estimated values of the model parameters
		are $ H_{0}=70.160, \alpha = 73.043 $ and $\beta =  1.223. $}
\end{figure}  
From the  Planck 2018 observation data, the observed value of the shift parameter is, $\mathcal{R}=1.7502 \pm 0.0046$ \cite{Shift2018} with $z_{2} = 1089.92.$ 

In the case of BAO, we followed the value of baryon acoustic oscillation peak from the SDSS data, $\mathcal{A} = =0.484 \pm 0.016$ having the sample redshift $ z_1 = 0.35 $\cite{Blake2011}. The theoretical value of the baryon acoustic peak \cite{EisensteinBAO2005} is,
\begin{equation}\label{eqn:A}
\mathcal{A}=\frac{\sqrt{\Omega_{m}}}{h(z_{1})^{\frac{1}{3}}}\left(\frac{1}{z_{1}}\int_{0}^{z_{1}}\frac{dz}{h(z)}\right)^{\frac{2}{3}}.
\end{equation}
We computed the $\chi^{2}$ using the standard relation, 
\begin{equation}
\chi^2(\alpha,\beta) = \sum \left[ \frac{A_i - A_i(\alpha,\beta,H_0)}{\sigma_i^2} \right], 
\end{equation}
where $A_i$ is the observed value of the cosmological parameter ( which can be $\mu_i, H_i, \mathcal{R}$ or $\mathcal{A}.$)  at a given redshift, $A_i(\alpha,\beta,H_0)$ is the theoretical value of the same parameter corresponding to the same redshift, and $\sigma_i$ is the uncertainty in the measurement of the
observable. 
Here $\mu_i$ is defined in equation (\ref{eqn:mui}), $H_i$ is the Hubble parameter at the respective redshifts in the Hubble parameter data set, $\mathcal{R}$ is as given in (\ref{eqn:R}) and $\mathcal{A}$ is as given (\ref{eqn:A}).
We have used the the 
	data set, SNe+CMB(Planck 2018)+BAO + HZ52, and extracted the parameters corresponding to the solution in equation (\ref{eqn:Hm1}) using Markov chain Monte Carlo (MCMC) method \cite{emsee2013}. The corner plots with marginal likelihood of the model parameters are in the figure \ref{fig:bestfit}. 
The best estimated parameters with 68.3 \% confidence level are summarised below in Table.\ref{table1}.
\begin{table}[h]\centering
	\begin{tabular}{ p{3.0cm}| c c c c }
		\hline \hline
		Data & $H_{0}$ &  $\alpha$ & $\beta$ & $\Omega_{\Lambda}$ \\
		\hline \\
		SNe Pantheon+\newline Planck (2018) + BAO+ HZ 52 &$70.160^{+0.687}_{-0.693}$ & $73.043^{+1.032}_{-1.019}$ & $ 1.223^{+0.001}_{-0.001} $ & $ 0.7246^{+0.0348}_{-0.0233} $ 
		\\
		\hline
	\end{tabular}\caption{The best obtained model parameters (with $ \chi^{2}_{min}=1069.531 $) using Markov chain Monte Carlo method } \label{table1}
\end{table}
The corresponding constraint for the age of the universe is found to be $H_0t_0 \sim 0.95630 - 1.0161,$ which is 
matching with the observational age as per the  WMAP and Planck observations, $H_0t_0 \sim 0.9644-1.0168$
\cite{wmap2003,wmap2006,wmap2008,wmap2010, planck2013,planck2015}.
The extracted value of $H_0t_0$ also agrees with the age constraint $H_0t_0 \geq \frac{1}{\beta^2} \sim  0.6674 - 0.6697 $ as per equation (\ref{eqn:ageconstrint}). It is also to be noted that the cosmological constant $\Omega_{\Lambda}$ extracted, is in agreement with a previous estimate given in reference \cite{Sahni:2002}.

\section{Interpreting the expansion of universe as emergence of cosmic space}
\label{sec:4}

This section describes how the solution obtained explains the evolution of the universe as the emergence of space. According to the emergent paradigm, the expansion of the universe is treated as the emergence of space, driven by the 
discrepancy between the DoF on the horizon and that have emerged within the volume bounded by the horizon.
This is what is depicted in equation (\ref{eq:law}). 

\begin{figure}[ht]
	\centering
	\includegraphics[width=0.45\textwidth]{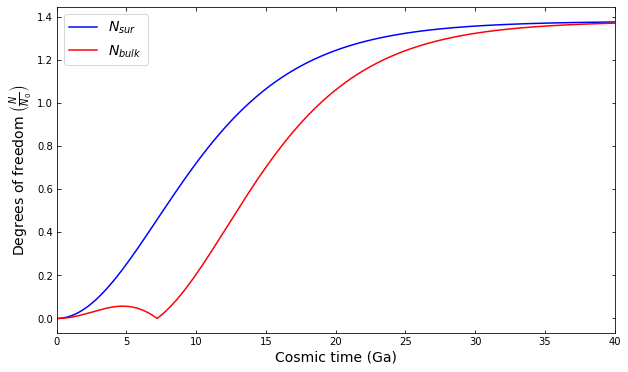}
	\caption{ \label{fig:newDoF} The Plot of evolution of DoF with the progress of cosmic time $t$ using the extracted model parameters. The top curve represents the evolution of $N_{sur}$ and the curve below represents the evolution of $N_{bulk}$ }
\end{figure}

Combining equations (\ref{eqn:DoF}), (\ref{eqn:Komarnew}) and (\ref{eqn:Hm1}),   the DoF on the horizon and that in the bulk can now be expressed as,
\begin{equation}\label{eq: Nsur new}
N_{sur} =\frac{4\pi c^{2}}{L_{P}^{2}}\frac{1}{H^{2}}=\frac{4\pi c^{2}}{L_{P}^{2}}\frac{\beta^{2}}{\alpha^{2}}\left(\frac{e^{2\alpha\beta t}-1}{e^{2\alpha\beta t}+1}\right)^{2},
\end{equation}
\begin{equation}
\label{eq: Nbulk new}
N_{bulk} =
\frac{4\epsilon\pi c^{2}}{L_{P}^{2}}\left[\frac{\beta^{4}}{\alpha^{2}}\left(\frac{e^{2\alpha\beta t}-1}{e^{2\alpha\beta t}+1}\right)^{4}+\frac{\beta^{2}\left(1-\beta^{2}\right)}{\alpha^{2}}\left(\frac{e^{2\alpha\beta t}-1}{e^{2\alpha\beta t}+1}\right)^{2}\right],
\end{equation}
where we have 
substituted for $B_1 \& B_2$ in terms of the model parameters $\beta$ and $\alpha$ respectively. The evolution of these two with the cosmic time is given in figure \ref{fig:newDoF}.  
It is evident from the plot, that the degrees of freedoms, $N_{sur}$ and $N_{bulk}$ evolve to decrease the discrepancy between them.  Finally, the discrepancy becomes zero, at which $dV/dt=0.$  
The Hubble parameter attains the constant value at this juncture, and 
expansion becomes de Sitter type.   

It is to be noted that, in figure (\ref{fig:newDoF}), there is a little hump in the early evolution of the bulk degrees of freedom. In the early stage,  $N_{bulk}$ increases first, attains a local maximum and then decreases to zero. After that, it increases and finally approaches $N_{sur}.$  This local hump is due to the combined effect of the evolutionary behaviours of the Komar energy density, $\rho+3P$ and $V/T,$ the ratio between the Hubble volume and the temperature. The factor $\rho+3P,$ which is positive in the very early stage of the universe, decreases to zero and then flips its sign to have negative values as the universe expands. Once $\rho+3P<0$, the universe begins to accelerate. However, $N_{bulk}$ is actually proportional to $|\rho+3P|$, which seems to decrease first, attains a minimum and then increases. Contrary to this, $V/T$ steadily increases from the beginning.  The combined effect of these behaviours is responsible for the  local hump in the evolution of $N_{bulk}.$ 
The end point of this local hump corresponds to the switch-over of the  expansion from the decelerated epoch into the accelerating epoch.

The DoF on the horizon, $ N_{sur} $ increases with the expansion and becomes a constant asymptotically. The important
point here is that $ N_{sur} $ and $ N_{bulk} $ attain saturation in the final stage. Secondly, these saturated values are equal such that $ N_{bulk} $ approach $ N_{sur} $ in the asymptotic limit. Since the emergence of space is driven by the discrepancy between $ N_{sur} $ and $ N_{bulk} $ ,
the asymptotic merging of these two implies the attainment of equilibrium. This is what essentially implied by the law of expansion.

\section{Conclusions} 
\label{sec:5}

The evolution of the universe can be explained as the emergence of space with the progress of cosmic time. According to this, the expansion of the universe can be described by the equation, $dV/dt = L_p^2 (N_{sur} - \epsilon N_{bulk} ),$ which explains that the evolution of the universe is driven by the difference between the degrees of freedom residing on the horizon and that within the bulk enclosed by the horizon.  In the present work we have derived a cosmological model from this principle. We assumed a suitable form for the   
Komar energy density as, $\rho+3P = -B_{2}+B_{1}H^{2} + B_{3}\dot{H},$ using which we have analytically solved the above equation of emergence, for Hubble parameter. 
The model describes a satisfactory evolution of the universe, which transit from a  
prior decelerated epoch to a late accelerated epoch.  
The model has been contrasted with the 
combined cosmological data, which includes Supernovae Type Ia, H-Z data, BAO and CMB \cite{Pantheon1048data,hz52data,Shift2018,Blake2011}, and the model parameters has been extracted.  

The evolution of the universe in this model is represented in figure \ref{fig:newDoF}. It shows that the degrees of freedom within the horizon, evolves as the universe expands and finally attains equilibrium with the degrees of freedom on the horizon. At this equilibrium stage, $N_{sur}$ and $N_{bulk}$ are equal, and the horizon becomes static without any further dynamical evolution.

At this juncture, we have to compare the present work with a previous one by Wang et al.\cite{wang1}, in which the authors assumed a constant  
Komar energy density, by retaining provision for varying density and pressure of the cosmic components. Even though the model described the evolution of the universe as the emergence of space, it could predict only the late accelerating epoch and failed to predict the transition from the prior decelerated to late accelerated epoch. In contrast, the present model predicts the transition from decelerated epoch to the late accelerated epoch. It also predicts more acceptable value for the cosmological constant. 

Finally, we would like to state that what we have proposed is not just an alternative model to cure any discrepancy of the existing cosmological models including $\Lambda$CDM. Instead, we highlighted that it is possible to derive a viable cosmological model from the principle of the emergence of space. 
It is to be mentioned that the principle of emergence has been extended to general theories of gravity like Gauss-Bonnet and more general Lovelock model \cite{cai,Sheykhi2013}. Also, the near-horizon field equations in these theories of gravity can be interpreted as a thermodynamic identity, and the underlying thermodynamic connections are well studied   \cite{PhysRevD.38.2434,AKBAR20067,Cai_Kim,Padmanabhan:2013,Chakraborty2015,Chakraborty2017}. Moreover, it has been shown that the equation of emergence can be derived by applying the first law of thermodynamics to the apparent horizon of the universe in the context of these higher-order gravity theories \cite{2018,Hareesh_2019}. These, in turn, imply that the emergence of cosmic space is very much related to horizon thermodynamics. In reference \cite{krishna2,krishnampla}, the authors have shown that the emergence of cosmic space can be treated as the quest of the universe to achieve a state of maximum entropy in Lovelock and other general gravity theories. As a future perspective, it will be interesting to investigate the extended versions of our model in the context of Gauss-Bonnet and Lovelock gravity theories by suitably defining the Komar energy density, possibly with higher-order corrections.   
\\[0.15in]

\noindent {\bf Acknowledgements}  
Hassan Basari V.T acknowledges Cochin University of Science and Technology for financial support. P. B. Krishna acknowledges KSCSTE, Govt. of Kerala, for the fellowship. \\

\noindent {\bf Data availability}
 
The pantheon data underlying this article are available at
\url{https://github.com/dscolnic/Pantheon} and is also available in \cite{Pantheon1048data}. The CMB data is availed from \cite{Shift2018}. The observational Hubble data (H-Z data) is availed from
\cite{hz52data}. The BAO data is availed from
\cite{Blake2011}.
\\


\begin{thebibliography}{99}	
	\bibitem{AKBAR20067}M. Akbar and Rong-Gen Cai,  \textit{Physics Letters B} \textbf{635}, 7-10 (2006).
	\bibitem{Cai_Kim} Cai, Rong-Gen and Kim, Sang Pyo, \textit{JHEP}, \textbf{02}, 50 (2005).
	\bibitem{paddy2010dec} T. Padmanabhan, {\it Mod. Phys. Lett. A} \textbf{25}, 1129 (2010).
	\bibitem{paddynew3} T. Padmanabhan, {\it Classical Quantum Gravity} \textbf{19}  5387 (2002). 
	\bibitem{paddynew4}  T. Padmanabhan, {\it AIP Conf. Proc.} 1241, 93 (2010).   
	\bibitem{paddynew1} T. Padmanabhan and A. Paranjape, {\it Phys. Rev. D}  \textbf{75}, 064004 (2007). 
	\bibitem{paddynew2} T. Padmanabhan, {\it Gen. Rel. Grav.} \textbf{40}, 529  (2008).
	\bibitem{sheykhi2007} Ahmad Sheykhi,Bin Wang and  Rong-Gen Cai,Phys. Rev. D \textbf{76}, 023515 (2007).
	\bibitem{disformalfr2019}Geng, Chao-Qiang and Hsu, Wei-Cheng and Lu, Jhih-Rong and Luo, Ling-Wei, \textit{Entropy} \textbf{21}, 172 (2019).

	\bibitem{bekenstein1} J. D. Bekenstein, {\it Phys. Rev. D} \textbf{7}, 2333 (1973).
	\bibitem{bardeen} J. M. Bardeen, B. Carter, S.W. Hawking, {\it Comm. Math. Phys.} \textbf{31}, 161 (1973).
	\bibitem{bekenstein2} J. D. Bekenstein, {\it Phys. Rev. D} \textbf{9}, 3292 (1974).
	\bibitem{Hawking1} S. W. Hawking, {\it Phys. Rev. D} \textbf{13}, 191 (1976).	
	\bibitem{Hawking2} S. W. Hawking, {\it Comm. Math. Phys.} \textbf{43}, 199 (1975).	
	\bibitem{1995} T. Jacobson, {\it Phys. Rev. Lett.}  \textbf{75}, 1260 (1995).

	\bibitem{Verlinde} E. Verlinde, {\it J. High Energy Phys.} \textbf{04}, 029 (2011).	
\bibitem{Kothawala2007} D. Kothawala et. al., {\it Physics Letters B} \textbf{652} (5), 338-342 (2007).
	\bibitem{Kothawalanew1} D. Kothawala and T. Padmanabhan, {\it Phys. Rev. D} \textbf{79}, 104020 (2009).
	\bibitem{Chakraborty2015}Sumanta Chakraborty and T. Padmanabhan, \textit{Phys. Rev. D} \textbf{92}(10), 104011 (2015).
	\bibitem{Chakraborty2017}Sumanta Chakraborty,\textit{ Classical and Quantum Aspects of Gravity in Relation to the Emergent Paradigm },Springer International Publishing (2017). 
	\bibitem{Paddy:2017intgratingconstant} T. Padmanabhan, {\it J. Phys. Conf. Ser.} \textbf{880}, 012008 (2017).			
	\bibitem{paddy2012jun} T. Padmanabhan, arXiv, arXiv:1206.4916 (2012).
	\bibitem{cai} R.G. Cai, {\it J. High Energy Phys.}  \textbf{2012}, 16 (2012).	
	\bibitem{Sheykhi2013} A. Sheykhi, {\it Phys. Rev. D}  \textbf{87}, 061501 (2013).
	\bibitem{hamsa} T. Padmanabhan and Hamsa Padmanabhan, {\it Int. J. Mod. Phys. D} \textbf{23} 1430011  (2014).
	\bibitem{paddycosmologicalconstant} T. Padmanabhan,  arXiv:1210.4174 (2012).
	\bibitem{paddyhamsacosmin} T. Padmanabhan, H. Padmanabhan, {\it Int. J. Mod. Phys. D} \textbf{22}, 1342001 (2013).
	
	\bibitem{krishnampla} P. B. Krishna and T. K. Mathew, {\it Mod. Phys. Lett. A}  \textbf{35},  2050334 (2020).	
	\bibitem{FLDezaki} F. L. Dezaki, B. Mirza, {\it Gen. Rel. Grav.} \textbf{47}, 67 (2015).	
	\bibitem{2018} M. Mahith, P. B. Krishna, T. K. Mathew, {\it J. Cosmol. Astropart. Phys.} \textbf{12}, 042 (2018).
	\bibitem{Hareesh_2019} T.  Hareesh, P. B. Krishna, T. K. Mathew, {\it J. Cosmol. Astropart. Phys.}  \textbf{12}, 024 (2019).
	\bibitem{krishna1} P. B. Krishna, T. K. Mathew, {\it Phys. Rev. D} \textbf{96}, 063513 (2017).
	\bibitem{krishna2} P. B. Krishna, T. K. Mathew, {\it Phys. Rev. D} \textbf{99}(2), 023535 (2019).
	\bibitem{Padmanabhan:2003Grentropy} T. Padmanabhan, {\it Class. Quant. Grav.} \textbf{21}, 4485 (2004).
	 \bibitem{Padmanabhan2014} T. Padmanabhan, \textit{Gen. Rel. Grav}, \textbf{46}, 1673 (2014).
	\bibitem{runningvacuum} J. Sola, A. Gomez-Valent,  {\it Int. J. Mod. Phys. D}  \textbf{24}, 1541003 (2015).	
	\bibitem{generalcovariance1} I. L. Shapiro, J. Sola, {\it J. High Energy Phys.}  \textbf{02}, 006 (2002).				
	\bibitem{generalcovariance2} I. L. Shapiro, J. Sola, {\it Nucl. Phys. Proc. Suppl.} \textbf{127}, 71 (2004).
	\bibitem{generalcovariance3} I. L. Shapiro, J. Sola,   {\it Adv. High Energy Phys.}  \textbf{2003} , 013 (2004).
	\bibitem{generalcovariance4} J. Sola, {\it J. Phys. A} \textbf{41}, 164066 (2008).
		
	\bibitem{phantomproblems1} S.M. Carroll, M. Hoffman, M and Trodden, Phys. Rev. D \textbf{68}, 023509 (2003).
	\bibitem{phantomproblems2} J.M. Cline, S. Jeon and G.D. Moore, Phys. Rev. D \textbf{70}, 043543 (2004).		
	\bibitem{Pavon2013} D. Pavon, N. Radicella, {\it Gen. Rel. Grav.} \textbf{45}(1), 63 (2013).
	\bibitem{Egan} C. A. Egan, C. H.  Lineweaver, {\it Astrophys. J.} \textbf{710}(2), 1825 (2010).
	\bibitem{Pavon1} N. Radicella, D. Pavon, {\it Phys. Rev. D} \textbf{89}(6), 067302 (2014).
	\bibitem{Pantheon1048data} Scolnic D. M. et al., {\it APJ}  \textbf{859}, 101 (2018). 

	\bibitem{hz52data}Geng, Jia-Jia and Guo, Rui-Yun and Wang, Anzhong and Zhang, Jing-Fei and Zhang, Xin,  {\it Commun. Theor. Phys.} \textbf{70},(4) 445 (2018).
	\bibitem{Shift2018}Yin, Zhao-Yu and Wei, Hao
	{\it Eur. Phys. J. C} \textbf{79}(8), 698 (2019).
	
	\bibitem{Blake2011} Blake C. et al.,{\it MNRAS} \textbf{418}, 1707 (2011).
	\bibitem{cmb1997} J. R. Bond, G. Efstathiou and  M. Tegmark, {\it Mon. Not. R. Astron. Soc.} \textbf{291}(3), L33 (1997).
	\bibitem{EisensteinBAO2005} D. J. Eisenstein, et. al.,  {\it Astrophys. J.}  \textbf{633}, 560 (2005).	
	\bibitem{emsee2013} D. J.{Foreman-Mackey}, Daniel and {Hogg}, David W. and {Lang}, Dustin and {Goodman}, Jonathan,  {\it Publ. Astron. Soc. Pac.}  \textbf{125}, 306 (2013).
	\bibitem{wmap2003} D. N. Spergel, et. al., {\it Astrophys. J. Suppl.} \textbf{148}, 175 (2003).								
	\bibitem{wmap2006} D. N. Spergel, et. al., {\it  Astrophys. J. Suppl.} \textbf{170}, 377 (2007).								
	\bibitem{wmap2008} G. Hinshaw, et. al., {\it Astrophys. J. Suppl.} \textbf{180}, 225 (2009).
	\bibitem{wmap2010} E. Komatsu,et. al.,  {\it Astrophysical J. Suppl} \textbf{192}(2), 18 (2011).
	\bibitem{planck2013} P. A. R. Ade, et. al.,  {\it Astron. Astrophys.} \textbf{571}, A1 (2014).
	\bibitem{planck2015} P. A. R. Ade, et. al.,  {\it Astron. Astrophys.} \textbf{594}, A13 (2016).	
	\bibitem{Sahni:2002} V. Sahni, T. D. Saini, A. A. Starobinsky and U. Alam, {\it JETP Lett.}  \textbf{77}, 201 (2003).	
	\bibitem{wang1} Z. L. Wang, W. Y.  Ai, H. Chen, J. B. Deng, {\it Phys. Rev. D} \textbf{92}, 024051 (2015).

\bibitem{PhysRevD.38.2434} Robert C. Myers and Jonathan Z. Simon, 
\textit{Phys. Rev. D} \textbf{38}, 2434 (1988).
\bibitem{Padmanabhan:2013} T. Padmanabhan and D. Kothawala, \textit{Phys. Rept.} \textbf{531}, 115--171 (2013).


						
\end{thebibliography}
\end{document}